# Properties of materials considered for improvised masks

**Steven. N. Rogak[1], Timothy A. Sipkens[1], Mang Guan[1], Hamed Nikookar[1], Daniela Vargas Figueroa[2], Jing Wang[3]**

[1] Department of Mechanical Engineering, University of British Columbia
[2] Forest Products Biotechnology, University of British Columbia
[3] Department of Anesthesia, Surrey Memorial Hospital, 13750 96 Avenue, Surrey, British Columbia, Canada V3V 1Z2

## ABSTRACT

During a pandemic in which aerosol and droplet transmission is possible, the demand for masks that meet medical or workplace standards can prevent most individuals or organizations from obtaining suitable protection. Cloth masks are widely believed to impede droplet and aerosol transmission but most are constructed from materials with unknown filtration efficiency, airflow resistance and water resistance. Further, there has been no clear guidance on the most important performance metrics for the materials used by the general public (as opposed to high-risk healthcare settings). Here we provide data on a range of common fabrics that might be used to construct masks. None of the materials were suitable for masks meeting the N95 NIOSH standard, but many could provide useful filtration (>90%) of 3 micron particles (a plausible challenge size for human generated aerosols), with low pressure drop. These were: nonwoven sterile wraps, dried baby wipes and some double-knit cotton materials. Decontamination of N95 masks using isopropyl alcohol produces the expected increase in particle penetration, but for 3 micron particles, filtration efficiency is still well above 95%. Tightly woven thin fabrics, despite having the visual appearance of a good particle barrier, had remarkably low filtration efficiency and high pressure drop. These differences in filtration performance can be partly explained by the material structure; the better structures expose individual fibers to the flow while the poor materials may have small fundamental fibers but these are in tightly bundled yarns. The fit and use of the whole mask are critical factors not addressed in this work. Despite the complexity of the design of a very good mask, it is clear that for the larger aerosol particles, *any* mask will provide substantial protection to the wearer and those around them.

## 1. INTRODUCTION

SARS-CoV-2, the virus responsible for the COVID-19 pandemic, has killed half a million people by July 2020 (World Health Organization, 2020). Disease transmission largely results from virus-containing particles[1] produced via coughing, sneezing, vocalizations and even normal breathing. While the degree of asymptomatic transmission is highly uncertain, with estimates in the range of 25-85% of cases (Mizumoto, et al., 2020; Li, et al., 2020), asymptomatic cases are likely to play some role in disease transmission. This highlights the need for public health measures that can be applied very widely to prevent widespread community transmission. Research is increasingly suggesting that masks, including non-medical masks, may be able to significantly reduce COVID-19 transmission (Eikenberry, et al., 2020; Stutt, et al., 2020; Lyu & Wehby, 2020; Leung, et al., 2020; Leffler, et al., 2020) and led the World Health Organization (WHO) and Center for Disease Control (CDC) to recommend universal facial covering and mask use for the general public. At the same time, the global reach of the pandemic saw a shortage of personal protective equipment for healthcare providers, requiring conservation of the supply of

[1] We will generally denote these suspended, virus-containing particles as aerosols unless there is a need to distinguish between fresh droplets or dried particles. Some of the challenges associated with this distinction for transmission of COVID-19 are discussed by Asadi et al. (2020), Morawska and Cao (2020), and some of the references therein.

N95 respirators and surgical masks for the healthcare providers. The result was a range of improvised masks, some of which were constructed with little guidance in terms of the effectiveness of materials.

While not as effective as surgical masks and respirators (MacIntyre, et al., 2015; Wilson, et al., 2020), cloth masks have a long history. Masks improvised from gauze and cotton wool were an important part of controlling the Manchurian plague (Lynteris, 2018). Research has demonstrated that cloth masks can protect others from the wearer (van der Sande, et al., 2008; Anfinrud, et al., 2020) – by reducing the total mass and volume of droplets relayed into the atmosphere and the distance that droplets travel (Dbouk & Drikakis, 2020; Viola, et al., 2020; Kumar, et al., 2020) – and offers some protection for the wearer (Chu, et al., 2020). While some of the effectiveness of non-medical masks depends on how they are worn, properly fit testing the population would suggest that improvised mask performance is largely determined by how well the masks are made. In response to the plethora of instructions on mask making on the internet and limited understanding of the factors controlling mask effectiveness, the WHO (World Health Organization, 2020), Institute of Medicine (2006), the National Academies of Sciences (Besser & Fischhoff, 2020), and, to a lesser extent, the Royal Society (2020) have published interim guidance on fabric mask use. These publications call for more research to be conducted in inform on how to make, fit, use and clean cloth masks.

To this end, the present manuscript provides measurements of various features relevant to improvised mask design, such as suitable filtration abilities, good fit and sufficient comfort (including breathability). Cloth masks used by the general public are also meant to be re-used, so they must be amenable to a practical decontamination process, such as laundering. We also consider the WHO interim guidance on fabric masks (World Health Organization, 2020) that recommends a 3-layer construction[2]: (*i*) an innermost layer to provide a comfortable biocompatible contact with the wearer and should readily absorb moisture produced be the wearer; (*ii*) a hydrophobic outer layer acts to limit surface contamination, repel the largest respiratory droplets incident on the mask, and act as an initial protection from the mask from getting wet[3]; and (*iii*) a middle layer provides most of the filtration. We primarily focus on filtration efficiency and breathability across a wide range of fabrics and filters, with an emphasis on biocompatible materials. Attention is placed on the materials, rather than the fit of the mask, which is another important feature in practice and motivates more breathable and comfortable materials. We aim to systematically characterize the fabric type (woven, woven brushed, knit, knit brushed, knit pile), material type (cotton, polyester, silk, wool, nylon, spandex, polypropylene), fabric parameters (fabric weight, thickness, water resistance) and construction type (number of layers) in order to offer rational performance metrics for the design of fabric masks that can guide policy makers.

## 2. BACKGROUND ON PARTICLE FILTRATION

### 2.1. *Respiratory particle size distribution*

Before discussing filtration, it useful to discuss the size range of the particles that need to be filtered. Respiratory particles are generated via different mechanisms involving the small airways (airway reopening mechanism), the vocal cords (physical vibrations, opening and closure) and the oral pharynx (dispersion of saliva between the epiglottis and lips) (Bake, et al., 2019). The particles are emitted as liquid drops but eventually dry as suspended aerosols. The resulting (wet) size distribution is very broad (0.01 micron to beyond 1mm mm), with multiple overlapping modes, including ones centered around 1-10 microns and around 100-200 microns (Johnson, et al., 2011), cf. Figure 1, with some variability between studies. For loud speech, the number and size of particles increase (Asadi, et al., 2019) to the point that energetic coughing can produce a very broad range of particles with a substantial fraction above 100 microns (Johnson, et al., 2011; Duguid, 1946; Xie, et al., 2007). The majority of the mass is contained in the larger *droplets*, which typically settle on the order of several meters of the source but could be carried further by winds or the momentum from a powerful sneeze (Bahl, et al., 2020; Nicas, et al., 2005; Bourouiba, 2020). Viral transmission through these large droplets could occur via (*i*) fomite transfer from contaminated surfaces onto mucosal surfaces or (*ii*) direct inhalation. While the former mechanism is dominant for large droplets, inhalation efficiencies of particles up to 100 micron may exceed 20% (Millage, et al., 2010). As such, there exists large uncertainties in the relative significance of these mechanisms as a function of particle size (Morawska, 2006; Judson & Munster, 2019; Fiegel, et al., 2006; Xie, et al., 2007; Chen & Zhao, 2010; Tellier, 2009; Leung, et al., 2020). Fortunately, almost any cloth mask will act to deflect droplet-laden jets during exhalation, which would effectively filter all particles larger than 10 microns on exhalation or inhalation and protect bystanders from an infected individual.

[2] While it could be possible to combine many of these functions in a mask with fewer (or more) than 3 layers, this guideline nevertheless provide a useful framework for evaluating materials.

[3] The role of the outer hydrophobic layer in a non-surgical mask is limited, as any filter layer will effectively block all but a powerful and persistent spray. In applications with such risks, the use of a face shield is more advisable.

On the other side of the size spectrum, submicron respiratory droplets carrying the virus[4] can accumulate, remain airborne, and contain viable SARS-CoV-2 virus (Fears, et al., 2020) for many hours. Emitted as liquid droplets, these particles would dry in seconds to solid particles of approximately half the diameter (note Figure 1 shows the wet aerosol size distributions). The production of these aerosols also varies widely between individuals (Fiegel, et al., 2006; Asadi, et al., 2019), leading to the phenomenon of *superemitters*. However, the drastic reduction in the volume will result in the number of viruses at these smaller sizes being orders of magnitude less than in particles of 1-2 microns[5]. This is convolved with the deposition fraction of particles in the human respiratory system, which shows a minimum of ~8% at 0.3 microns, before increasing to ~30% at 1 micron and 100% at 10 microns (Park & Wexler, 2008). Thus, while it is well established that the submicron particle filtration efficiency of cloth materials is very low (0.7 to 50%) (Rengasamy, et al., 2010; Konda, et al., 2020; Jayaraman & et al., 2012), the risk posed by these smallest particles is likely quite small. Accordingly, we propose that designs should focus on particles in the 1-5 micron size range and, while we consider a larger range of sizes, discussion will be centered about this size range. It is worth noting that the NIOSH N95 tests use a sodium chloride aerosol with an aerodynamic diameter below this range and will typically feature higher penetration rates.

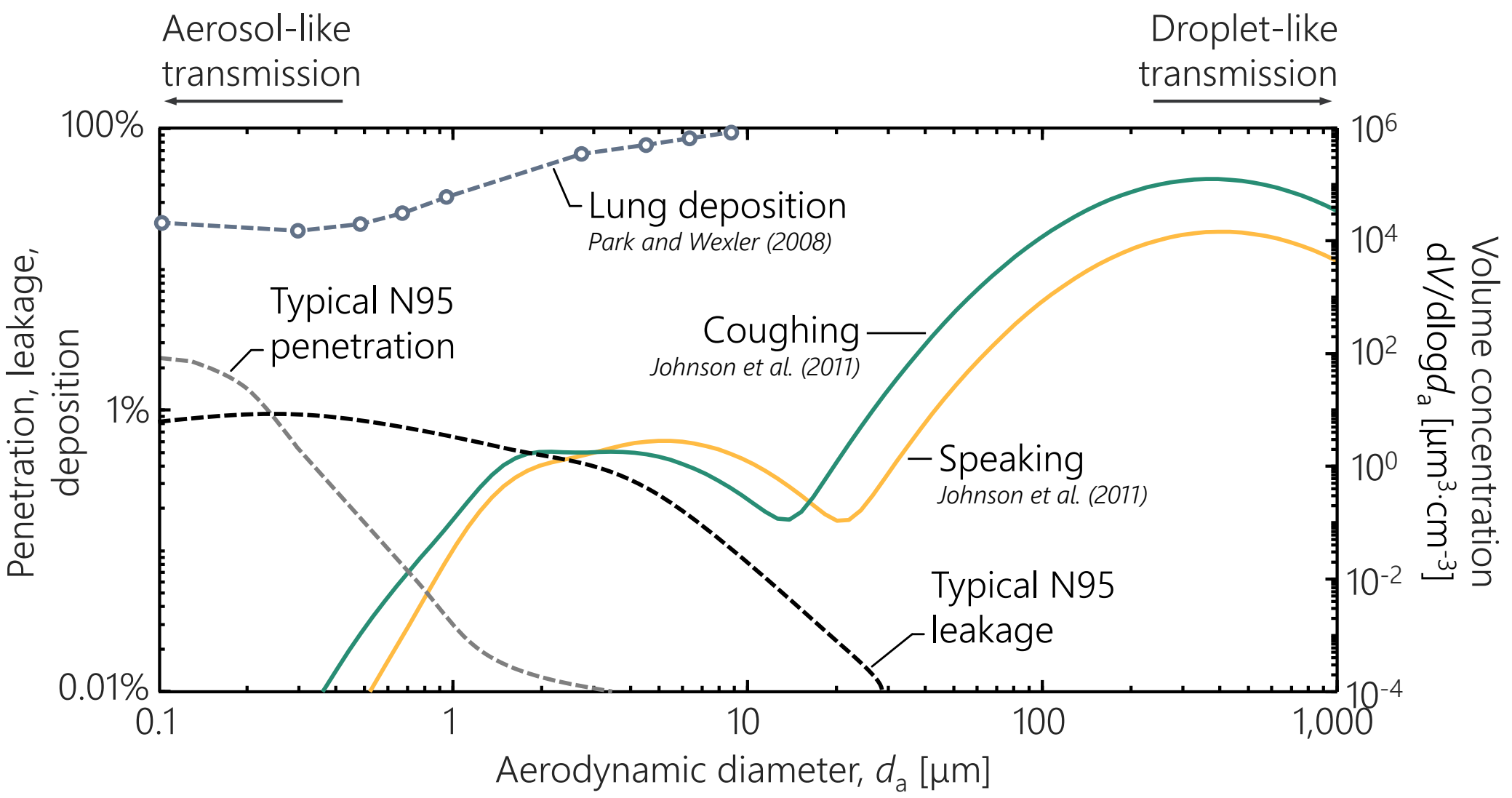

**Figure 1.** Size distributions and penetration rates relevant to face masks. Solid lines correspond to volume concentrations on the right axis, while dashed lines correspond to percentages on the left axis. Coughing and speaking bioaerosol distributions are taken from the BLO, tri-modal model of Johnson et al. (2011). See Nicas et al. (2005) for a summary of other, older measurements of these distributions. Lung deposition fraction is taken from Park and Wexler (2008). Typical N95 curves are approximate and are compiled from multiple sources, including measurements by the authors and those by Huang et al. (2007). While not a discrete change, larger particles are more likely to settle on surfaces, resulting in droplet-like transmission, while smaller particles are likely to stay aerosolized, resulting in aerosol-like transmission.

[4] SARS-CoV-2 has a diameter of 0.06 to 0.14 microns (Zhu, et al., 2020; Kim, et al., 2020), representing a hard minimum in terms of particle size.

[5] Milton et al. (2013) affirmed that influenza virus is present in particles below 5 micron but only presented a coarse particle size binning of greater or less than 5 micron.

## 2.2. Mask design and key material properties

Particle filtration is governed by four main mechanisms. For the smallest particles (< 0.1 micron), deposition in the filter is controlled by Brownian motion, and hence the mobility-equivalent diameter of the particles. For larger particles (> 0.5 microns), deposition is controlled by impaction on mask fibers, such that the aerodynamic diameter is the most useful measure of size. The interception mechanism becomes more important for particles comparable the filter fiber size. Filtration of particles in the 0.1 to 2 micron range can also be strongly influenced by electrostatic forces. Most or all N95 masks rely on the manufactured charges of electret media, which is why N95 tests require neutralization of the challenge aerosol and why certain decontamination methods (isopropyl alcohol, laundering) can greatly reduce the performance of N95 masks. The relative importance of these different mechanisms is influenced by face velocity (i.e., air flow divided by filter surface area) effects, where low velocities allow more time for aerosol particle removal by electrostatic attraction, which decreases particle penetration, and high velocities make inertial impaction more effective.

The overall performance of a particle filter can be assessed using the size-dependent penetration, $P(d_p)$ which is the percentage of particles that penetrate through the mask,

$$P(d_p) = \frac{N_{in}}{N_{out}}, \tag{1}$$

where $N_{in}$ and $N_{out}$ are the concentrations of particles of diameter $d_p$ inside and outside of the mask, respectively. The filtration efficiency is

$$\eta = 1 - P. \tag{2}$$

Thus, an N95 mask must have an efficiency $\eta > 0.95$ or a penetration $P < 0.05$, which will apply for the specific particle characteristics and flow conditions of a given test[6]. Arbitrarily high filtration efficiency could be obtained with any material if enough layers are used, but this could result in an unacceptable flow resistance as well as unacceptable thickness or weight. The tradeoff between filtration efficiency and airflow resistance is instead commonly characterized by the quality factor $Q$ (Podgorski, et al., 2006; World Health Organization, 2020; Zhao, et al., 2020)[7],

$$Q = -\ln(P)/\Delta p = -\ln(1-\eta)/\Delta p, \tag{3}$$

where $\Delta p$ is the pressure drop for the face velocity used in the test. A useful property of $Q$ is that samples with $N$ identical layers should have approximately the same value of $Q$, stemming from multiplicative penetrations and additive pressure drops. Table S1 in the Supplemental Information lists select values of $Q$; their $\log_{10}$ equivalents, which would correspond to those reported by Zhao et al. (2020); and the expected filtration provided specific values of pressure drop (e.g., the penetration for a given *Q* and a 30 Pa pressure drop), which will be referenced in the discussion.

[6] Different standards apply to surgical mask (ASTM Bacterial filtration efficiency, FDA particle filtration efficiency) and respirator (NIOSH particle filtration efficiency) filter testing in North America. In terms of size, tthe ASTM Bacterial Filtration Efficiency test uses aerosolized Staphylococcus Aureus (0.6-0.8microns), with a relatively large mean aerodynamic diameter of 3 microns.

[7] The literature varies in whether $\ln(\cdot)$ or $\log_{10}(\cdot)$ is used. We use $\ln(\cdot)$, consistent with Podgorski et al. (2006), which results in larger quality factors than those presented in some works, including Zhao et al. (2020).

Reported filtration efficiency by a given study depends on the air face velocity, degree of filter loading, aerosol size and aerosol charge distribution. These parameters are not always completely specified and complicate direct comparison between different masks and respirators (Rengasamy, et al., 2012; Rengasamy, et al., 2017). In filter testing, it is possible to use a monodisperse challenge aerosol, but more often the mask is challenged with an aerosol having a broad size distribution and concentrations are measured as a single integral over size. In the NIOSH N95 test, for example, the NaCl challenge aerosol has a geometric mean size of 0.075 micron and geometric standard deviation of ~1.8. However, photometric measurements of concentration weight the larger sizes more heavily, so that the effective challenge size is roughly 0.2 micron (considering optical properties of NaCl and typical photometric instruments). Further, since the density of NaCl is larger than that of water, a 0.2 micron physical diameter results in an even larger aerodynamic diameter around 0.3 microns. In general, the lack of standardized filter testing poses a problem for the interpretation of both peer-reviewed studies and the growing body of non-peer reviewed pandemic research on improvised mask materials. Various face velocities (corresponding to flows of 15-116 L/min), aerosol sizes (0.1-10 micron) and aerosol types (NaCl aerosols, ambient particles, Bacillus arophaeus, Bacteriophage MS2, Staphylococcus Aureus) have been reported or left unspecified (Rengasamy, et al., 2010; Jayaraman & et al., 2012; Davies, et al., 2013; Jung, et al., 2013; Konda, et al., 2020; Bagheri, et al., 2020; Agency for Science, Technology, and Research, 2020; Mueller & Fernandez, 2020; Schempf, 2020).

The filtration efficiency desired for a mask material depends indirectly on mask fit. The actual protection provided by a respirator depends also on leakage between the mask and the face. Fit testing typically reports a *fit factor* which is the inverse of the apparent penetration, including leakage, and is more variable in that it depends on the fit, breathing frequency and other characteristics of individual persons (He, et al., 2014). Thus, if the challenge aerosol and flow conditions are the same in the material and the fit testing, leakage would necessarily result in apparent penetrations that are larger than in material testing (such as the NIOSH N95 test). However, the test conditions are generally not the same for material and fit testing. In fit testing, typically only charged particles are sampled, resulting in extremely high filtration efficiency such that the recorded apparent penetration is a measure of leakage only. A properly fitted N95 mask will have a fit factor over 100, while a typical, poorly-fitted surgical mask will have a measured fit factor < 3, despite similar filtration properties (Derrick & Gomersall, 2005). However, with improper fit, the performance of N95 masks also drops to that of surgical masks (Grinshpun, et al., 2009) such that fit testing is crucial to the proper use of N95 respirators for frontline workers at increased risk of exposure to viral aerosols. Because poor fit will likely dominate particle exposure when wearing improvised masks, there would be little benefit in constructing such masks of high efficiency filter media.

# 3. METHODS

## *3.1. Challenge aerosol*

In the present work, a sodium chloride aerosol was generated from solution with an ultrasonic mesh nebulizer (Sonar MedPro). Initial tests were done at 15 g/L concentration, but this was increased to 20 g/L to improve particle counting statistics at the larger sizes[8]. The output of the nebulizer was diluted with room air in an extraction duct, resulting in 50 ±5% RH and concentrations below 3000 #/cc. Room temperature was 21 ±2°C for all tests. The portion of the aerosol used for the filter testing was passed through an x-ray neutralizer (TSI model 3088); this does not result in an uncharged aerosol but rather one with a quasi-equilibrium bipolar charge distribution.

[8] For approximately the last 10% of tests, which focused on the effect of washing and drying, aerosols were generated using at TSI 3076 Constant Output atomizer. This was more stable but required salt concentrations approaching 80 g/L.

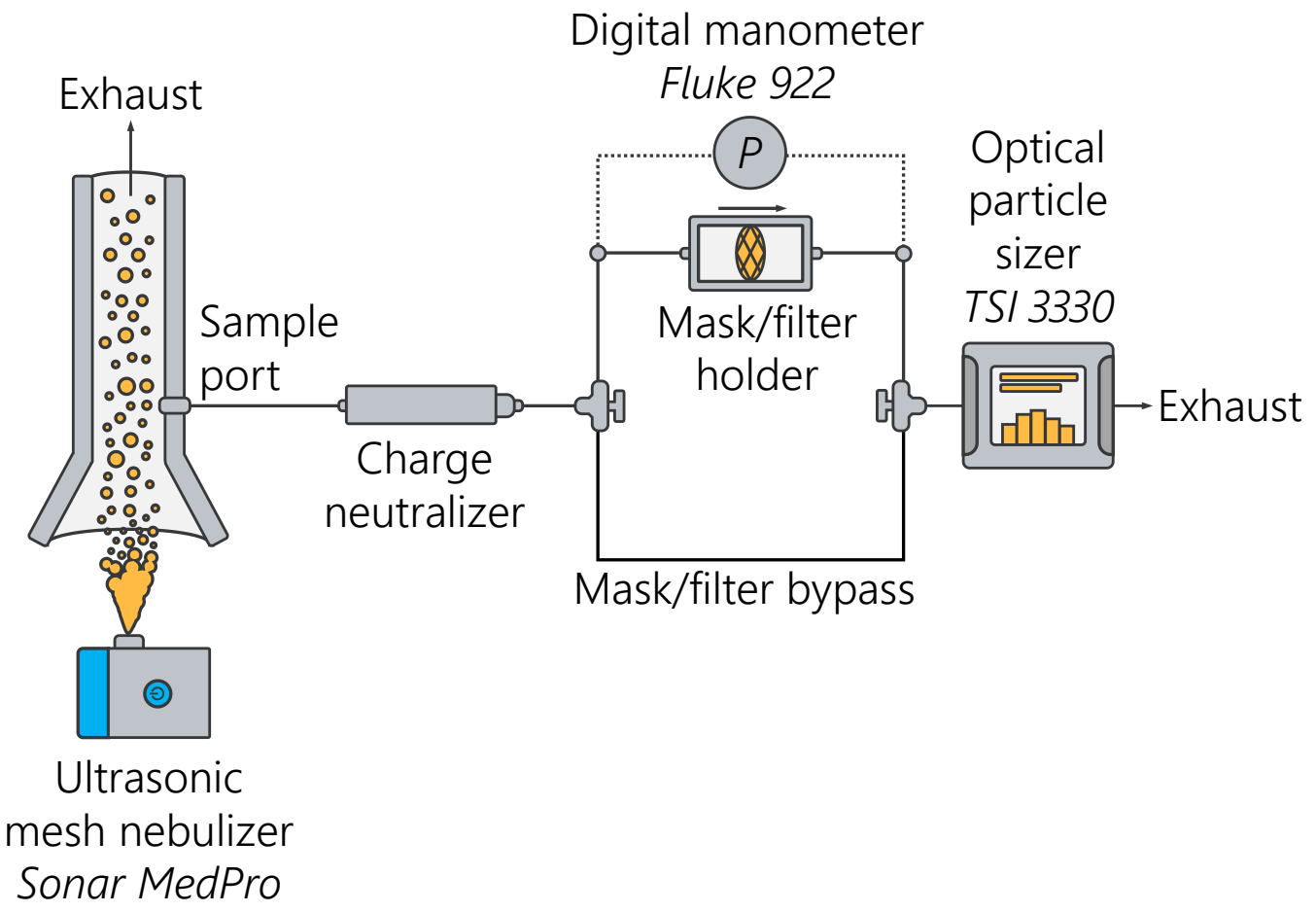

**Figure 2.** Schematic of the experimental setup used in testing the face mask materials. The line lengths for the mask holder and its bypass are the same length, to avoid the effect of line losses.

### *3.2. Apparatus*

After charge neutralization, the aerosol flows through the tested filter punch, or a bypass. Pressure differential was measured using a Fluke 922 digital manometer. The flow rate through the filter is controlled by the downstream instruments. For most of the tests, this flow was 1.0 LPM (set by the downstream OPS), resulting in a face velocity of 4.9 cm/s through the sample. This is less than the velocity in the NIOSH N95 test (8-9 cm/s); at lower flow rates, deposition by electrostatics would be increased, but deposition by impaction would be decreased. Rengasamy (2010) used challenge particles up to 1 micron and found that the effect of face velocity on the filtration efficiency of common fabrics was quite modest, but at 1 micron, lower face velocity results in *higher* penetration – presumably due to the importance of impaction. Thus, we expect that our reported results are conservative in the sense that higher breathing rates would result in greater capture of the larger aerosols.

The bypass line and main filter path have the same length and are made from stainless steel tubing 0.375” OD. The flow path (bypass or filter) is controlled by 2 three-way valves, symmetrically arranged so that losses in bends are identical for both paths.

From the filter (or the bypass), flow was sent to an Optical Particle Sizer (OPS TSI 3330). All penetration measurements involved a sequence of measurements of the bypass, then the sample, then the bypass. Samples during the transition between sample/bypass were discarded, and the penetration was based on the ratio of counts of particles penetrating the sample divided by the average of the two bypass periods. The OPS was run with 16 channels over the maximum range allowed. The bin limits are defined using the default calibration (for PSL particles) and are reported in the Table S2 in the Supplemental Information.

The bins from the default OPS procedure correspond to scattering cross-sections, σ, which are then corrected for refractive index (1.4 assumed here for NaCl) to estimate the physical size of the salt particles at the bin limits. The representative physical size, $d_g$, is taken as the geometric mean of the upper and lower bounds. Finally, this geometric size is converted to aerodynamic diameter, which is more often the quantity of interest for inhalation and filtration, using

$$d_a = d_g \left[\frac{\rho_p C_c(d_g)}{\rho_0 C_c(d_a)}\right]^{1/2}, \tag{4}$$

where $d_a$ is the aerodynamic diameter, $\rho_p$ = 1 g/cm$^3$ is the unit density, $\rho_p$ is the density of the particle material, and $C_c$ is the Cunningham correction factor. This conversion, for each channel, is also indicated in Table S1 in the Supplemental Information. To simplify screening of materials, we primarily consider the penetration at Channel 9, corresponding to a physical salt particle size of 2 microns and an aerodynamic size of 2.8 microns – which is near the mass median size of the fine mode of respiratory particles (cf. Figure 1). Size-dependent penetrations are shown in the results for selected materials.

### 3.3. Material characterization

A major challenge in evaluating cloth masks is that materials are typically supplied without detailed characterization. Although fabric manufacturers may have information on the material blends, thread counts, weights and weave, this information is not available at the retail level, and might vary from batch to batch in consumer products. Thus, it is challenging to specify materials in this study so that results can be reproduced by scientists or individuals making masks could be assured of the effectiveness reported here. We have partly mitigated this problem by including materials of well-defined products, and by including basic physical characterization of the materials.

**Table 1.** Description of material codes along with their descriptions, sub-categories and full set of examples.

| Code | Fabric structure | Material | Examples |
|---|---|---|---|
| K | Knit | Cotton | **K1** ( Single knit jersey, cream); **K2** ( Single knit jersey, grey); **K3** ( Ribbed knit cotton ); **K4** ( Double knit jersey, yellow ) |
| | | Cotton blend | **K5** ( Fine gauge, single knit jersey, beige ); **K6** ( Single loopback knit cotton); **K7** ( Double knit jersey, salmon); **K8** (Fine-gauge, single knit jersey, 5% lycra); **K9** ( Single knit jersey, 5% lycra) |
| | | Spandex blend | **K10** (Spandex polyester); **K11** (Nylon) |
| W | Woven | Cotton-based | **W1** (Gauze); **W2** (Batik cotton); **W3** (Downproof cotton); **W4** (Flannel); **W5** (Quilting cotton); **W6** (Cotton, 600TC) |
| | | Cotton spandex | **W7** (Spandex cotton); **W8** (Spandex cotton) |
| | | Polyester (*includes W17*) | **W9** (Polyester satin); **W10** (Polyester peel ply); **W11** (Polyester); **W12** (Polycotton); **W13** (Spandex PC); **W17** (Chiffon) |
| | | Wool | **W14** (Wool blend); **W15** (Melton wool) |
| | | Silk | **W16** (Silk) |
| nW | nonwoven | Polypropylene, **nW2-4** (Halyard) | **nW1** (Interfacing polypropylene); **nW2** (H300); **nW3** (H400); **nW4** (H600); **nW5** (Dried baby wipe) |
| | | Cellulose | **nW6** (Commercial washroom towel); **nW7** (Paper towel) |
| | | Microfiber | **nW8** (Microfiber) |
| CP | Cut pile | Velour | **CP1** (Velour) |
| | | Polyester | **CP2** (Fleece); **CP3** (Velvet); **CP4** (Corduroy) |
| - | Multilayer masks | - | **ASTM2** (Surgical mask); **N95** (3M 1860); **NMM** (Non-medical mask); **W16+W11** (Silk + Polyester) |

Sample weight was determined using a milligram balance, and sample area was determined using calipers. Cloth weave and fiber diameter were determined for select materials by optical microscopy. A qualitative measure of hydrophobicity was obtained using a custom surrogate for the standard textile spray test (International Organization for Standardization, 2017), where select materials here were subjected to a gravity-fed spray of 500 mL of water. Samples were placed at a 45° angle to the vertical spray, such that water was allowed to run off of the material during testing. Samples were placed at a distance of approximately 15 cm from the nozzle. The change in weight before and after the test was measured using a microbalance. Effectively, this is a measure of the material's ability to wick moisture, rather than spray blocking efficiency. However, Adyin et al. (2020) tested fabrics for droplet blocking efficiency and found that 2 or 3 layers of even the most porous materials (ie tee-shirt) are very effective at blocking large droplets- thus droplet blocking efficiency is not a significant issue for the multilayer masks.

### 3.4. Material selection and experimental design

Forty-one individual materials were tested in addition to a reference surgical mask (ASTM Level 2 Primagard), respirators (N95, 3M 1860), and commercial non-surgical mask (NMM, Greenlife). The full list of materials is provided in nonwoven fibrous materials that create tortuous paths with air pockets to achieve high filtration efficiency and breathability (wool felt, fiberglass paper, and polypropylene). Based on these principles, we screened fabrics with structures composed of looping and interlocking layers (knitted pile, double knit, gauze) and agitated surface fibers (wool, brushed fabric), some of which demonstrated promising filter qualities in previous research (Jayaraman & et al., 2012). Although the variations of materials within each of these categories are large, we have selected a few examples to illustrate some of the most important differences between material types.

Table 1, with more detail provided in Table S3 and a data file in the Supplemental Material. The samples fall into five broad categories according to the structure of the fabrics: woven (W), knit (K), cut pile (velour, velvet, fleece, corduroy on a woven or knit base), nonwoven (spunbond, spunlace, paper), and multilayer (i.e., full mask structures). The majority of clothing fabrics were cotton and polyester, while the majority of the nonwoven materials were polypropylene and cellulose. Traditional filters of surgical masks and respirators contain layers of

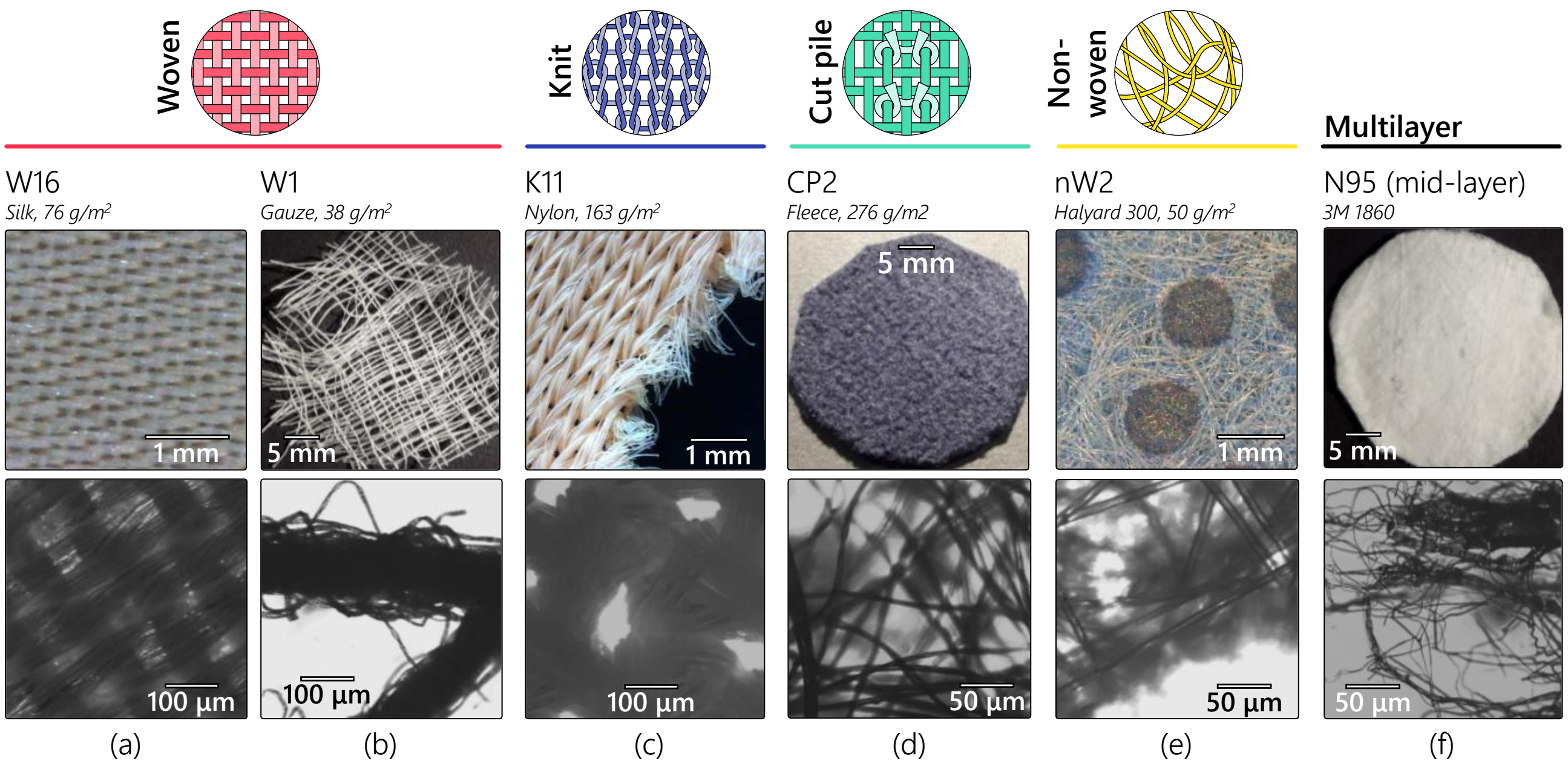

**Figure 3.** Optical microscopy images of various materials considered in this work, each at two magnifications. The higher magnification images have a consistent scale in (*a*-*c*) and in (*d*-*f*). Selected materials include examples of two woven materials using natural fibers, silk (W16) and gauze (W1); synthetic fabrics with a knit base, nylon-spandex knit (K11) and polyester fleece (CP2); a N95 mask; and the Halyard 300 material (nW2). Scanning electron microscopy by Zhao et al. (2020) can act to supplement these observations, e.g., indicating the mat-like microstructure typically formed by the cellulose materials.

Woven fabrics contain yarns (bundles of fibers) that are interlaced at right angles to each other. Figure 3 shows fabrics with (a) a tight weave (natural silk) and (b) an extremely loose weave (cotton medical gauze) and are generally consistent with the scanning electron microscopy (SEM) images of a woven cotton or nylon by Zhao et al. (2020). Although the silk fibers are quite thin (under 10 microns), the yarns are tightly bundled, and an aerosol particle travelling through the fabric would interact mainly with yarns rather than individual fibers. The medical gauze is so open that any mask composed of this material would use many layers, effectively creating a random matrix of the yarns. In this case, the yarns are fibrillated such that particles would have significant interactions with individual fibers which are below 10 microns in diameter. Other cotton materials exhibited such fibrillated yarns. We screened certain material types like polyester and silk that tend to retain static discharge, which have been found to improve filtration through postulated electrostatic interactions (Konda, et al., 2020).

Knit fabrics use more complicated interlacing of bent (often looping) yarns, and are usually more stretchable than woven fabrics. Figure 3 also shows (c) a tightly knitted nylon-spandex material and (d) polyester fleece. The fleece has a knit base but is categorized here as *cut pile* because the thick fuzzy layer anchored to the knit would presumably have a large influence on filtration properties. In the cut pile layer of the fleece (this appeared similar to the SEM images of the polyester toddler wrap material in Zhao et al. (2020)), it appears that individual fibers are largely separated and thus the fiber (rather than yarn) dimension would control interactions with aerosol particles.

The category of nonwoven fabrics includes any sheet material formed from a random mat of fibers. This therefore includes papers made from cellulose fibers and mats of (commonly) polypropylene fibers that are produced by meltblowing then bonded either thermally (spunbond) or mechanically (spunlace). The synthetic plastic fibers of spun-bonded and melt-blown polypropylene in medical masks and respirators is capable of holding strong electrostatic charge that improve filtration at the submicron level, although the charge may dissipate with prolonged exposure to humidified air (Institute of Medicine 2006). We also screened biocompatible and potentially electrostatically charged polypropylene such as dried baby wipes (spunlace) and medical sterilization wrappings (spunbond and meltblown) that had been identified as effective improvised limited-use filter material from recent pandemic research (Bagheri, et al., 2020; Meijer & Vrielink, 2020). The manufacturer specification for the Halyard brand of sterilization wrapping claims that the fabric has been infused with electric micro-fields that surround the meltblown fibres to form charged gradients within the fabric. Figure 3 compares (e) a commercial sterile wrap material, Halyard 300 (nW2, bottom panel), which is also a spunbond, meltblown polypropylene, and (f) the middle layer of an N95 surgical mask (3M 1860). Both of these materials are electret (ie., contain semi-permanent electric charges on the surface).

## 4. RESULTS AND DISCUSSION

### *4.1. Filter layer and particle penetration*

The primary function of the mask is to filter particles while allowing the wearer to breathe. All of the layers of the mask contribute something to the filtration and the flow resistance. Materials were located on a plot of penetration for 2.8 micron particles vs pressure drop in Figure 4, where the size of the symbol indicates the material weight per unit area and isolines of quality factor are given in gray. The optimal filter material would occur at in the upper left region of the plot. Uncertainties are larger for the low filtration efficiency materials in the lower region of the panel, where the change in aerosol numbers is small relative to the overall aerosol concentration. Materials with high $Q$ (i.e., low pressure drop and high filtration) are automatically contenders for the middle filter layer of a mask. Contenders for the inner and outer layers must either have high $Q$ or very low single-layer pressure drop. Although there are no standards for the flow resistance of cloth masks, a natural target is 30 Pa (at 4.9 cm/s), roughly matching surgical masks (ASTM2). N95 masks can afford larger flow resistance because they are intended to be fit tested – a mask with high airflow resistance and poor fit would result in relatively large leakage around the mask. Assuming that the non-filtering layers of the mask should use a small portion of the pressure *budget*, only single layers with $\Delta p$ < 10 Pa appear viable as inner or outer layers, unless they also have high $Q$.

The N95 masks (3M 1860) and ASTM Level 2 Surgical mask have ~100% filtration efficiency at 2.8 microns (infinite $Q$). Even an N95 mask treated with IPA has much higher $Q$ than any of the cloth samples tested. The Halyard materials (nW2-4), dried baby wipes (nW5), and some knit (K) cotton materials have $Q$ in the range of 30-100 $kPa^{-1}$ and should be considered viable candidates for the filter layer of cloth masks.

The woven fabrics typically had a low-quality factor. Cotton gauze ($Q$ = 56 $kPa^{-1}$) was the notable exception. However, for a mask with 30 Pa pressure drop at the test conditions, one would need over 100 layers and the weight would be 4.5 $kg/m^2$ or 70 g for a typical mask area of 150 $cm^2$. Thus, while this gauze mask would remove over 80% of 2.8 micron particles, it would have a weight comparable to that of an elastomeric half-face respirator and over 20x higher than weight of a typical disposable face mask. A few other woven fabrics approached $Q$ = 30 $kPa^{-1}$ (flannel, W4; the wool samples, W14-15; and polyester chiffon, W17); where compiling enough layers to give a 30 Pa resistance, filtration efficiency would be ~60% at 2.8 microns (cf. Table S1). Woven silk (W9) had a relatively low quality factor ($Q$ ~ 15 $kPa^{-1}$), consistent with Zhao et al. (2020) (though that work focussed on submicron particles, which makes direct comparison challenging). Some of the other woven fabrics (quilting cotton, W5; polyester satin, W9; silk, W16; and polyester chiffon, W17) had low flow resistance that could be suitable for the non-filtering layers of a composite mask. The polyester peel ply material (used for manufacturing resin reinforced plastics) is included here as an extreme example of a tight weave. It is remarkable here because it had one of the highest pressure drops *and* the highest particle penetration of any material tested.

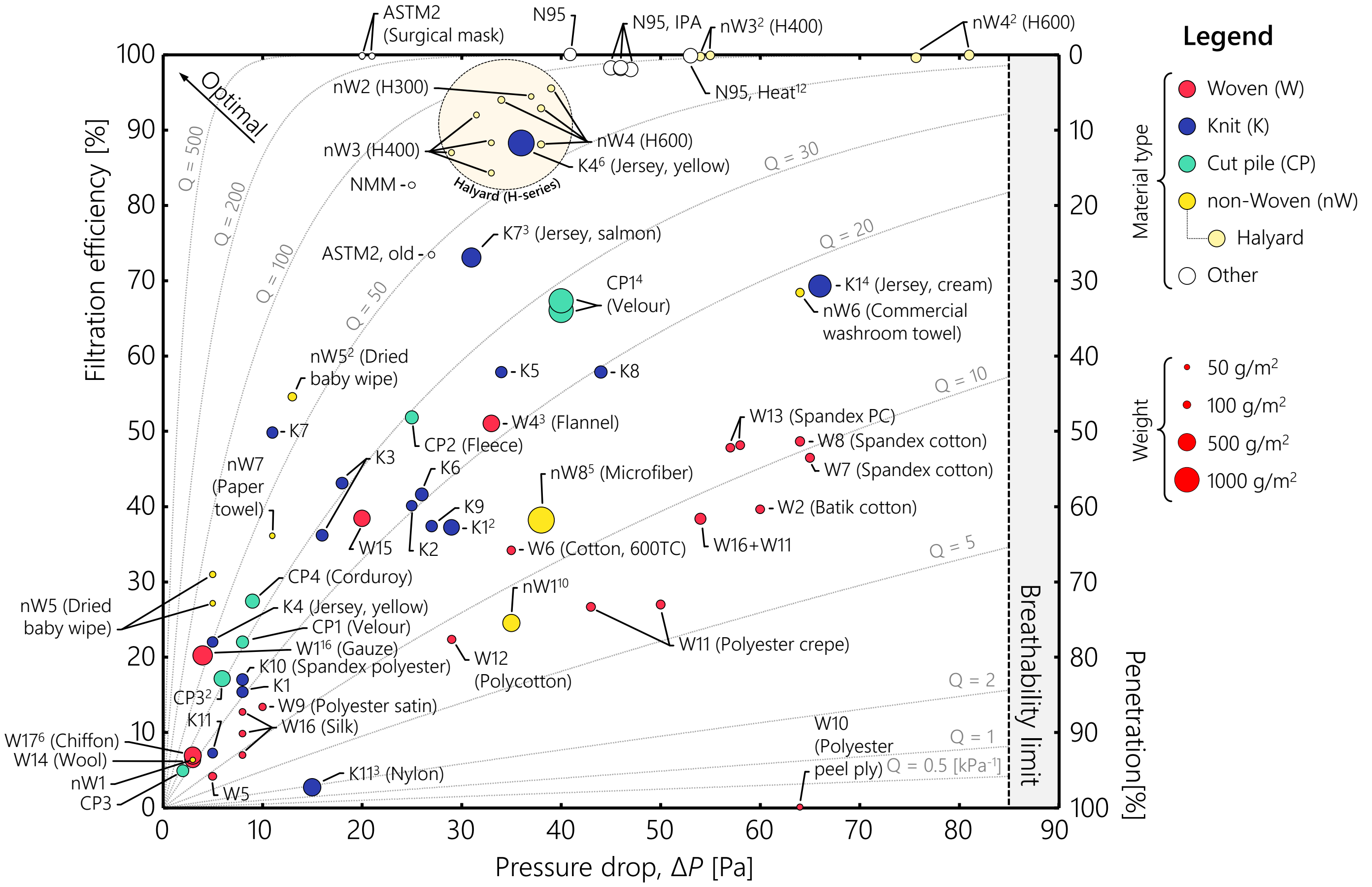

**Figure 4.** Pressure drop and material penetration for 2.75 micron aerodynamic diameter sodium chloride particles, a face velocity of 4.9 cm/s and a range of common materials. Superscripts, if present, refer to the number of layers of material used in the test (e.g., W1[16] is 16 layers of gauze). The 100% cotton down proof ticking occurs off of the plot in terms of pressure drop. The region containing the repeats for the single layers of the Halyard material is highlighted in the upper region of the plot. Full material names are given where space is allowed, otherwise material codes are only provided. Quality factor isolines are labelled in $kPa^{-1}$. An analogous plot for 1.9 micron aerodynamic diameter is shown in Figure S1 in the Supplemental Information.

The knitted fabrics tested had $Q$ > 9 $kPa^{-1}$, with several knitted cottons (Jersey, yellow, K4; Cotton jersey blend, dense, salmon, K7) having $Q$ > 50 $kPa^{-1}$. Used in a mask with 30 Pa flow resistance, $Q$ > 50 $kPa^{-1}$ corresponds to removing > 78% of 2.8 micron particles. The corresponding mask weight would be 10-20 grams.

Interestingly, the cut pile materials were all closely clustered about $Q$ ~ 30 $kPa^{-1}$. While fleece (CP2) could be used as a lightweight but moderately effective filter layer, the cut pile materials were typically much thicker than the plain woven or knit fabrics, rendering them less attractive for masks (on the order of 4 layers would be needed as the filter layer for a composite mask). The synthetic cut pile materials would be poorly suited to the inner (moisture wicking) layer or the outer layer, which one might wish to clean by wiping.

The nonwoven fabrics are by far the best candidates for a filter layer, often with $Q$ in the range of 30-100 $kPa^{-1}$ and with low weight. The efficiency of these fabrics is partly attributable to the fact that the fundamental fibers, though not necessarily thinner than those of the fabrics described earlier, are not bundled into yarns. Rather, the fibers are all fully exposed to the oncoming aerosol stream. In addition, the Halyard series of sterile wrappings ($Q$ > 50 $kPa^{-1}$) appear to be electret, given their response to alcohol and washing. The most promising materials (sterile wrap, dried baby wipes, and 2-ply paper towel) differ substantially in their mechanical and material properties, however. The Halyard sterile wrap is a strong and hydrophobic material that would be suitable for the reusable outer structure of mask. However, it is not biodegradable and washing with soap and water or cleaning with isopropyl alcohol consistently reduce $Q$ by almost a factor of 2 – bringing it down to the level of the dried baby wipes, some paper towels and the best of the knitted cotton fabrics. The dried baby wipes are widely available, washable without degradation, but not easily constructed into a mask and are not biodegradable. Interestingly, the favorable quality factors for dried baby wipes are shared with the work of Bagheri et al. (2020), where different types of baby wipes consistently had pressure drops below 10 Pa and filtration efficiencies above 50% (this was for a similar particle size, though the type of particle size, e.g., aerodynamic diameter, was not specified and for a face velocity of 7.3 cm/s). Quilted paper towel (Bounty 2-ply, nW7) could be used as a cheap and biodegradable filter layer that would contribute to the moisture removal from the inner layer. It is worth noting that coarse Kimberly Clark commercial washroom towels (nW6) did not perform as well quilted paper towels ($Q$ = 18).

Figure 5a shows size-resolved penetration curves for a range of materials, typically selecting some of the best and worst performing materials in each category. To fairly compare the filtration, we first corrected the penetrations to a sample thickness corresponding to a pressure drop of 30 Pa at 4.9 cm/s, assuming that $Q$ is independent of the number of layers used (the validity of this assumption is discussed later). This results in a corrected penetration of the form,

$$P_{\mathrm{corr}} = P^{30/\Delta p}, \qquad (5)$$

where $\Delta p$ is given in Pa. As noted previously, the woven materials tended to perform poorly, which was true over a range of particle sizes. Also consistent with previous observations (where all of the cut pile materials had similar quality factors), the penetration curves span a relatively narrow range. The knit fabrics often perform as well as the nonwoven materials across a range of particle sizes, though the nonwovens are often much lighter. The non-medical disposable mask (NMM) has filtration efficiency comparable to the best fabrics but would be an order of magnitude lighter.

Selected materials were tested before and after different cleaning treatments (cf. Table S4); the impact of these cleaning processes on the filtration was typically small or inconsistent except for some of the nonwoven fabrics, as noted below.

## 4.2. Inner Layer

A range of cotton, cotton-blends and silk fabrics were screened as potential materials for the innermost hydrophilic layer that should readily absorb moisture. Some of these materials were disqualified for having a low value of $Q$ and a high pressure drop (Batik cotton, W2; downproof ticking, W3; cotton-spandex weaves, W7-8; and the cotton-lycra knit, K8). All of these samples absorbed a substantial amount of water but not all would be suitable as an inner layer.

We tested the hydrophilicity of the three samples with the lowest airflow resistance: low thread-count woven cotton (quilting cotton, W5), double-knitted cotton (K7) and silk (W16). Double-knitted cotton (Δ260%) demonstrated the highest water absorbency, followed by silk (Δ210%) and quilting cotton (Δ140%). Overall, double-knitted cotton demonstrated the best characteristics for the innermost mask layer with low pressure drop ($\Delta p$ = 6-10 Pa), high quality factor ($Q$ = 63 kPa$^{-1}$), and high-water absorbency.

## 4.3. Outer Layer

We screened fifteen samples of knitted, woven and nonwoven materials for water resistance, pressure drop and particle penetration. The samples included the sterile wraps (nW3-5), woven polyester (W9, W11) and blends of nylon and spandex (K10-11). Brushed and cut pile fabric types were excluded as the entangled surface fibers tend to attract dust and other environmental contaminants. The 100% polyester lining and woven spandex blend samples demonstrated poor breathability (50-65 Pa) and were precluded from further consideration. H300-600 (Δ20-50%) and nylon (Δ70%) performed similarly to the surgical mask (Δ50%) for water resistance. Polyester satin (Δ100%) and polyester cotton (Δ130%) were more water resistant than woven and knit cotton (Δ140-260%), silk (Δ210%) and polyester spandex (Δ230%). Overall, nylon (knitted) and polyester satin (woven) demonstrated the best characteristics for the outermost mask layer with low pressure drop ($\Delta p$ = 5-10 Pa) and moderate hydrophobicity.

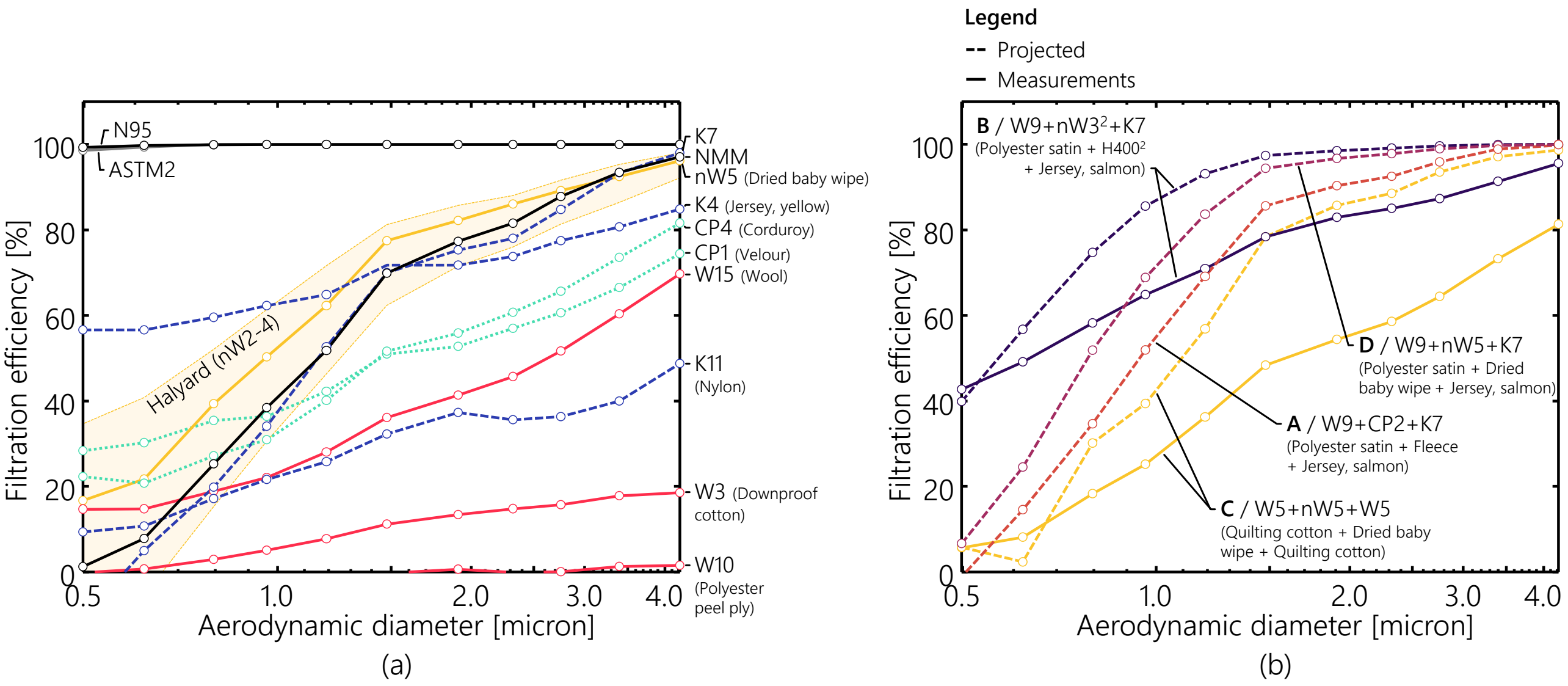

**Figure 5.** Filtration efficiency as a function of aerodynamic diameter. Within this range, smaller sizes are consistently more penetrative. Surgical and N95 masks have very low penetrations that are nearly coincident with the upper axis. (*a*) Results for single layer materials, normalized to 30 Pa of pressure drop across all materials. Materials are sampled in an attempt to span the range of observed penetrations for each fabric type (e.g., the woven materials generally have lower penetrations). (*b*) Filtration efficiency of multilayered candidate masks, where solid lines are measurements of specific material combinations and dashed lines correspond to projected penetrations (the product over multiple layers).

### 4.4. Multilayer Mask Performance

Using performance information for individual layers, it is possible to select materials for a multilayer mask. Here we verify the performance of selected material combinations and compare the results with masks manufactured to different standards.

In Figure 5b, we examine the filtration performance of fabric and filter hybrids using the 3-layer approach of combining a hydrophilic inner layer with a high filtration middle layer and a hydrophobic outer layer. Masks combinations achieved filtration efficiencies that approximated the single-use non-surgical mask (NMM) from a local pharmacy. Mask B using a double layer of Halyard sterilization wrapping as the filter achieved higher than or similar filtration efficiencies than the non-surgical mask. However, all of the non-medical masks (including the non-surgical mask) exhibited poor filtration efficiencies at the submicron range, in contrast to the surgical mask and the N95 respirator which achieved $\eta > 95\%$ across the entire particle size range of 0.5 to 10 microns[9]. Using the single-layer screening measurements for pressure drop and penetration, we have identified fabric mask combinations (Mask A-D) that demonstrate significantly improved filtration performance compared to the status quo, which typically comprise of 2 layers of woven cotton and a disposable filter like paper towel or dried baby wipe (Mask E).

The single-layer data can be used to estimate (dashed lines, Figure 5b) the performance of the multilayer mask assuming that the layers behave independently. However, this overestimates the efficiency of the multilayer mask substantially for 1-3 micron particles. This could be a result of the first layer preferentially removing charged particles, leaving the downstream layers to remove nearly uncharged particles. The average number of charges increases with particle size, but at a sufficiently large size, impaction becomes more important than electrostatic interactions.

Modern manufacturing technologies have taken advantage of the versatility of polypropylene to produce spun-bonded filters with fiber thickness down to micron or submicron diameters that are both lightweight and highly efficient at particle filtration. The surgical mask and N95 respirator we tested achieved $Q$ > 150 kPa$^{-1}$, which significantly outperformed the improvised filters we identified in this study ($Q$ = 30-100 kPa$^{-1}$). While Figure 5 demonstrates that effective fabric masks can be constructed with improvised materials, the thickness and the pressure drop are both greater for these masks compared to the single-use surgical and non-surgical masks. In some cases, the pressure drop, a measure of airflow resistance, approaches or exceeds that of N95 masks. The increased thickness of fabric masks can cause overheating and moisture buildup, while the increased airflow resistance can compromise breathability, both of which can negatively impact mask comfort and the frequency and safety of mask use. Therefore, mask designs using improvised materials should consider maximizing the surface area of air exchange to improve effective ventilation. A cup-shape or duck bill design may be preferred to the flat pleated design of surgical masks, which can often touch the face and not engage the entire surface area of the mask in air exchange.

---

[9] Indeed, measurements using a scanning mobility particle sizer by the authors revealed $\eta > 99\%$ for the N95 masks in the 0.02-0.3 micron mobility diameter range.

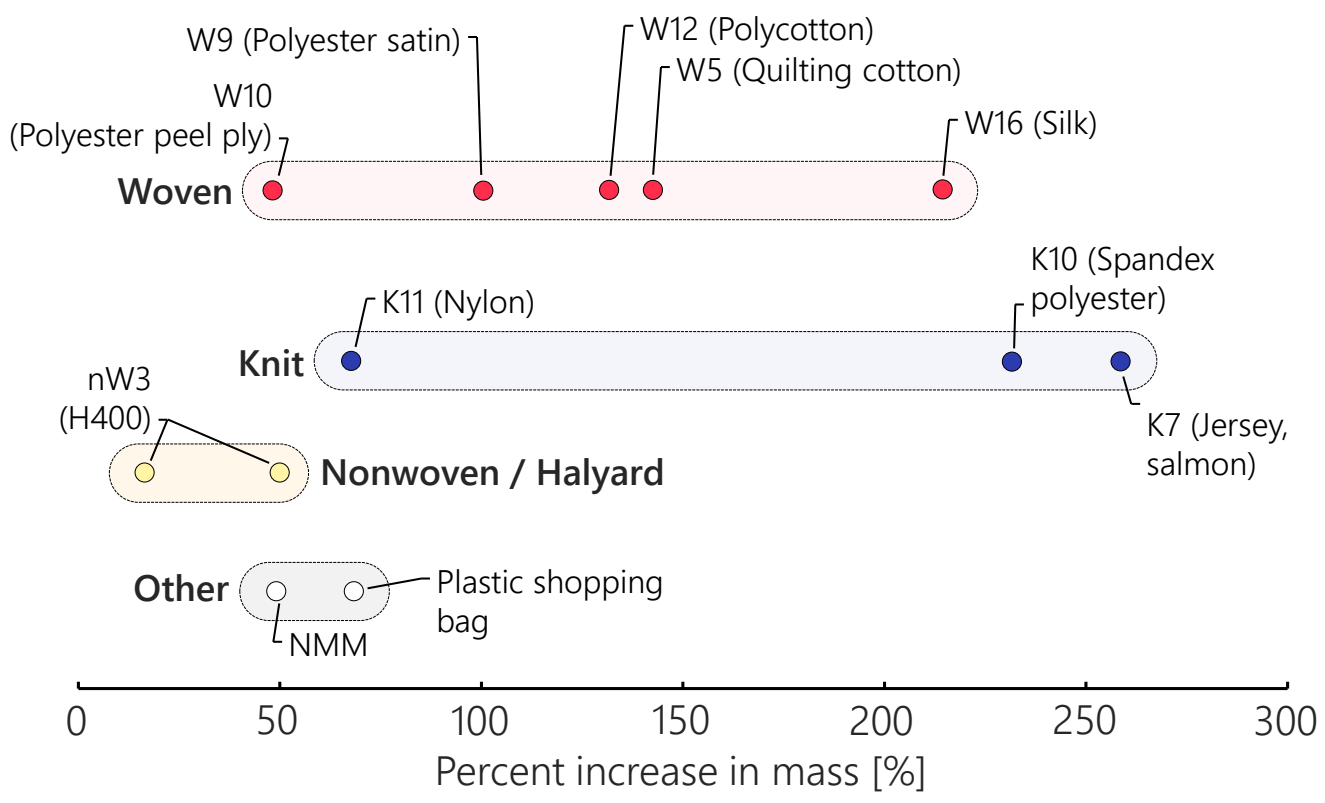

**Figure 6.** Percent increase in mass of samples following wetting with water. A standard plastic bag is shown for reference. The Halyard material and non-medical mask resulted in visible beading of water.

## 5. CONCLUSIONS

While common fabrics have poor filtration efficiency for submicron particles relative to N95 masks (even an N95 mask that has been washed in isopropanol has far better efficiency than fabrics over all particle sizes), most common fabrics are expected to be effective in removing large (~10 micron particles). Larger differences in the quality factor, $Q$, exists between the commercially available fabrics occur in the 1-5 micron range. Generally, tightly woven thin materials perform very poorly, and microscopy indicates that these materials form sheets with perforations as opposed to a matrix of thin fibers. Such materials with particularly poor filtration properties ($Q <$ 6 kPa$^{-1}$) include several spandex blends, silk and polyester satins. Looser weaves or knits of cotton perform relatively well ($Q \sim 50$ kPa$^{-1}$), possibly because the cotton yarns often exhibit a frayed surface, with smaller (~ 10 micron) fibers protruding from the main bundles. Interestingly, cotton gauze, the material used in the Manchurian plague masks, has a high quality factor but in practice requires a far heavier and thicker mask than the better of the modern alternatives.

Several nonwoven materials appear promising. The Halyard 300, 400 and 600 sterile wrappings (nW2-4) and dried baby wipes (which were also of interest in Bagheri et al. (2020)) are the best materials tested ($Q \sim 70$ kPa$^{-1}$) and perform only slightly worse than an isopropanol-washed 3M 1860 mask ($Q \sim 90$ kPa$^{-1}$). The H300, 400 and 600 samples were sensitive to washing with soap and water (or isopropanol), suggesting that these sterile wrapping materials rely substantially on electret properties. This would complicate cleaning and re-use of these masks, and forms of heat sterilization, developed previously for N95 masks, might be considered for this material. Dried baby wipes had a lower initial value of $Q$ but were not degraded by washing. Double ply paper towel would not be washable, but it is cheap, biodegradable, and has higher $Q$ than most fabrics. These lifecycle issues are not trivial given that billions of people will be wearing a mask daily during the COVID-19 (and potentially future) pandemic.

All of the fabrics tested had much lower $Q$ than commercial masks, which means that thicker masks would be needed to obtain good filtration. However, this should be balanced with the higher airflow resistance for thicker masks, which will result in more leakage around the mask when a proper seal is infeasible (fit testing the general public is infeasible, such that some leakage should be expected and low airflow resistance becomes an important design parameter). This implies the need for designs that have an appropriate shape and maximize the flow area of the mask by keeping a large portion of the cloth area off of the face. This in turn has implications for the stretchiness and stiffness of the fabric, which is beyond the scope of this paper. The inner layer of the mask should be biocompatible, soft, and wick water away from the face; cotton knits appear to be most suitable for this purpose. Three-layer masks provide the opportunity to use replaceable middle layers that might have high $Q$ but are not wicking, not washable and are less biocompatible. In such a 3-layer construction it is not entirely clear what characteristics are needed for the outer layer. WHO guidance is to use a hydrophobic layer to stop a liquid spray, but it is very hard to imagine that this is a serious concern for non-medical situations.

The size-dependent particle penetrations for all materials tested decreased monotonically and smoothly for all materials tested, thus there appears little opportunity for synergistic combinations of materials. Penetrations for multilayer combinations are noticeably higher than the product of the individual layer combinations, possibly due to electrostatic effects. Although it is challenging to use common fabrics to remove particles smaller than several microns, nearly all materials can remove most particles above 5 microns, and this supports the popular viewpoint that any mask is better than no mask in protecting the mask wearer *and* the people around them.

## FUNDING

This work has been funded by the Natural Science and Engineering Research Council of Canada (PDF-516743-2018) and the Canadian Council for the Arts (Postdoctoral Fellowship).

## ORCID

Steven N. Rogak · https://orcid.org/0000-0002-4418-517X
Timothy A. Sipkens · https://orcid.org/0000-0003-1719-7105

# Properties of materials considered for improvised masks

**Steven. N. Rogak[1], Timothy A. Sipkens[1], Mang Guan[1], Hamed Nikookar[1], Daniela Vargas Figueroa[2], Jing Wang[3]**

[1] Department of Mechanical Engineering, University of British Columbia
[2] Forest Products Biotechnology, University of British Columbia
[3] Department of Anesthesia, Surrey Memorial Hospital, 13750 96 Avenue, Surrey, British Columbia, Canada V3V 1Z2

## SUPPLEMENTAL INFORMATION

**Table S1.** List of quality factors, including the value of *Q* if $\log_{10}$ was used in the place of the natural logarithm in Eq. (3), and the expected penetrations given that the number of layers of a material was chosen to give a set pressure drop.

| | | **Filtration efficiency, η** | | | |
|---|---|---|---|---|---|
| ***Q*, ln [kPa$^{-1}$]** | ***Q*, log$_{10}$ [kPa$^{-1}$]** | $\Delta p$ = 5 Pa | $\Delta p$ = 10 Pa | $\Delta p$ = 20 Pa | $\Delta p$ = 30 Pa |
| 0.1 | 0.0434 | 0.050% | 0.10% | 0.20% | 0.30% |
| 0.2 | 0.0869 | 0.10% | 0.20% | 0.40% | 0.60% |
| 0.5 | 0.217 | 0.25% | 0.50% | 1.0% | 1.5% |
| 1 | 0.434 | 0.50% | 1.0% | 2.0% | 3.0% |
| 2 | 0.869 | 1.0% | 2.0% | 3.9% | 5.8% |
| 5 | 2.17 | 2.5% | 4.9% | 9.5% | 14% |
| 10 | 4.34 | 4.9% | 9.5% | 18% | 26% |
| 20 | 8.69 | 9.5% | 18% | 33% | 45% |
| 30 | 13.0 | 14% | 26% | 45% | 59% |
| 50 | 21.7 | 22% | 39% | 63% | 78% |
| 100 | 43.4 | 39% | 63% | 86% | 95% |
| 200 | 86.9 | 63% | 86% | 98% | 99.8% |
| 500 | 217 | 92% | 99.3% | 99.995% | 99.99997% |

**Table S2.** Optical particle sizer (OPS) channels and equivalent particles sizes for sodium chloride challenge aerosol. Bottom row indicates sizes corresponding to the mean diameter for standard NIOSH N95 challenge aerosol. All particle sizes are given in microns.

| | **Optical diameters** | | | |
|---|---|---|---|---|
| **OPS Channel** | Nominal bin lower limit | Refractive index-corrected lower limit | Geometric mean as bin center, $d_g$ | **Aerodynamic diameter, $d_a$** |
| 1 | 0.30 | 0.306 | 0.34 | 0.498 |
| 2 | 0.37 | 0.381 | 0.43 | 0.620 |
| 3 | 0.46 | 0.484 | 0.54 | 0.796 |
| 4 | 0.57 | 0.596 | 0.68 | 0.962 |
| 5 | 0.71 | 0.769 | 0.84 | 1.19 |
| 6 | 0.88 | 0.923 | 1.05 | 1.48 |
| 7 | 1.09 | 1.20 | 1.36 | 1.91 |
| 8 | 1.35 | 1.55 | 1.66 | 2.32 |
| 9 | 1.68 | 1.78 | 1.98 | 2.76 |
| 10 | 2.08 | 2.20 | 2.44 | 3.40 |
| 11 | 2.58 | 2.72 | 3.04 | 4.22 |
| 12 | 3.20 | 3.40 | 3.78 | 5.25 |
| 13 | 3.96 | 4.21 | 4.69 | 6.49 |
| 14 | 4.92 | 5.22 | 5.86 | 8.12 |
| 15 | 6.10 | 6.59 | 7.23 | 10.0 |
| 16 | 7.56 | 7.95 | 8.84 | 12.2 |
| *NIOSH N95* | - | - | *0.075* | *~ 0.3* |

**Table S3.** Full list of tested materials, including properties and select filtration efficiencies. Materials marked with a "*" represent the average over multiple repeats and standard deviations across the repeats are given for the filtration efficiency. Note, that this often involved only a single repeat, such that bounds on the filtration efficiency should be taken with some reservation (e.g., uncertainties are likely larger than that reported for the Polyester, W11, material). Filtration efficiency and quality factor are given for Channel 9, corresponding to an Aerodynamic diameter $d_a$ = 2.76 micron. Additional filtrations, repeats, and cases considering multiple layers of single materials are included in a separately piece of Supplemental Information.

| Code | Description | Structure | Material | Pressure drop, Δp [Pa] | Weight [g·m$^{-2}$] | Thickness [mm] | Filtration efficiency, η | Quality factor, $Q$ [kPa$^{-1}$] |
|---|---|---|---|---|---|---|---|---|
| W1 | Gauze | Woven | Cotton-based | 4 | 609 | 4.00 | 79.8% | 56 |
| W2 | Batik cotton | Woven | Cotton-based | 60 | 123 | 0.30 | 60.4% | 8 |
| W3 | Downproof cotton | Woven | Cotton-based | 169 | 144 | - | 38.2% | 6 |
| W4[3] | Flannel | Woven | Cotton-based | 33 | 453 | - | 49.0% | 22 |
| W5 | Quilting cotton | Woven | Cotton-based | 5 | 105 | 0.26 | 95.8% | 9 |
| W6 | Cotton, 600TC | Woven | Cotton-based | 35 | 121 | 0.25 | 65.8% | 12 |
| W7 | Spandex cotton, thick | Woven | Cotton spandex | 65 | 140 | - | 53.5% | 10 |
| W8 | Spandex cotton, thin | Woven | Cotton spandex | 64 | 144 | 0.27 | 51.3% | 10 |
| W9 | Polyester satin | Woven | Polyester | 10 | 86 | 0.23 | 86.6% | 14 |
| W10 | Polyester peel ply | Woven | Polyester | 64 | 65 | 0.07 | 99.9% | 0 |
| W11* | Polyester crepe | Woven | Polyester | 47 | 130 | - | 73.2 ±0.2% | 7 |
| W12 | Polycotton | Woven | Polyester | 29 | 112 | - | 77.6% | 9 |
| W13* | Spandex PC | Woven | Polyester | 58 | 125 | 0.23 | 52.0 ±0.3% | 11 |
| W14 | Wool blend | Woven | Wool | 3 | 417 | 2.20 | 93.6% | 22 |
| W15 | Melton wool | Woven | Wool | 20 | 440 | - | 61.6% | 24 |
| W16* | Silk | Woven | Silk | 8 | 76 | 0.18 | 90.1 ±2.9% | 13 |
| W17[6] | Chiffon | Woven | Polyester | 3 | 486 | - | 93.1% | 24 |
| K1 | Single knit jersey, cream | Knit | Cotton | 8 | 203 | 0.54 | 84.6% | 21 |
| K2 | Single knit jersey, grey | Knit | Cotton | 25 | 183 | 0.60 | 59.9% | 21 |
| K3* | Ribbed knit cotton | Knit | Cotton | 17 | 222 | 0.90 | 60.3 ±4.9% | 30 |
| K4 | Double knit jersey, yellow | Knit | Cotton | 5 | 182 | - | 78.0% | 50 |
| K5 | Fine gauge, single knit jersey, beige | Knit | Cotton blend | 34 | 212 | 0.60 | 42.1% | 25 |
| K6 | Single loopback knit cotton | Knit | Cotton blend | 26 | 270 | 0.82 | 58.4% | 21 |
| K7 | Double knit jersey, salmon | Knit | Cotton blend | 11 | 204 | 0.56 | 50.2% | 63 |
| K8 | Fine-gauge, single knit jersey, 5% lycra | Knit | Cotton blend | 44 | 254 | 0.70 | 42.1% | 20 |
| K9 | Single knit jersey, 5% lycra | Knit | Cotton blend | 27 | 215 | 0.59 | 62.6% | 17 |
| K10 | Spandex polyester | Knit | Spandex blend | 8 | 230 | 0.75 | 83.0% | 23 |
| K11 | Nylon | Knit | Spandex blend | 5 | 161 | 0.40 | 92.7% | 15 |
| CP1 | Velour | Cut pile | Velour | 8 | 242 | - | 78.0% | 31 |
| CP2 | Fleece | Cut pile | Polyester | 25 | 276 | 1.60 | 48.1% | 29 |
| CP3 | Velvet | Cut pile | Polyester | 2 | 224 | 0.70 | 95.1% | 25 |
| CP4 | Corduroy | Cut pile | Polyester | 9 | 331 | 1.40 | 72.6% | 36 |
| nW1 | Interfacing polypropylene | Nonwoven | Polypropylene | 3 | 49 | - | 93.6% | 22 |
| nW2 | H300 | Nonwoven | Polypropylene, Halyard | 37 | 50 | - | 5.5% | 78 |
| nW3* | H400 | Nonwoven | Polypropylene, Halyard | 32 | 57 | 0.40 | 12.1 ±3.2% | 67 |
| nW4* | H600 | Nonwoven | Polypropylene, Halyard | 37 | 82 | 0.53 | 7.4 ±3.2% | 70 |
| nW5* | Dried baby wipe | Nonwoven | Polypropylene | 5 | 59 | 0.41 | 70.9 ±2.7% | 69 |
| nW6 | Commercial washroom towel | Nonwoven | Cellulose | 64 | 118 | - | 31.6% | 18 |
| nW7 | Paper towel | Nonwoven | Cellulose | 11 | 53 | - | 63.9% | 41 |
| nW8[5] | Microfiber | Nonwoven | Microfiber | 38 | 1111 | - | 61.8% | 13 |
| ASTM2* | Surgical mask | - | - | 2 | 68 | 0.34 | 0.0% | 383 |
| N95 | N95, 3M 1860 | - | - | 41 | 335 | - | 0.0% | ~∞ |
| NMM | Non-medical mask | - | - | 25 | 71 | 0.45 | 17.3% | 70 |
| W16, W11 | Silk + Polyester | - | - | 54 | 205 | - | 61.6% | 9 |

**Table S4.** Select filtration efficiencies for masks that were sanitized with a variety of treatments, including laundering, heat treatments, cleaning with isopropanol (IPA), and cleaning with soap and water (SW). Repeat experiments are listed as separate entries.

| | | | | Filtration efficiency, η | | | Quality factor, $Q$ [$kPa^{-1}$] | | |
|---|---|---|---|---|---|---|---|---|---|
| **Code** | **Material** | **Treatment** | **Pressure drop, Δ*p* [Pa]** | $d_a$ = 0.962 | $d_a$ = 2.76 | $d_a$ = 5.25 | $d_a$ = 0.962 | $d_a$ = 2.76 | $d_a$ = 5.25 |
| W5 | Quilting cotton | - | 5 | 101.7% | 95.8% | 84.1% | 0 | 9 | 35 |
| | | Launder | 10 | 95.8% | 82.2% | 57.1% | 4 | 20 | 56 |
| W9 | Polyester satin | - | 10 | 98.3% | 86.6% | 68.6% | 2 | 14 | 38 |
| | | Launder | 5 | 93.5% | 87.8% | 73.0% | 13 | 26 | 63 |
| W12 | Polycotton | - | 29 | 93.6% | 77.6% | 59.0% | 2 | 9 | 18 |
| | | Launder | 24 | 93.1% | 74.7% | 51.8% | 3 | 12 | 27 |
| W15 | Melton wool | - | 20 | 84.7% | 61.6% | 33.1% | 8 | 24 | 55 |
| | | Launder | 25 | 83.9% | 63.1% | 40.0% | 7 | 18 | 37 |
| W16 | Silk | - | 8 | 97.2% | 90.1% | 84.9% | 4 | 13 | 20 |
| | | Launder | 7 | 92.6% | 81.2% | 72.8% | 11 | 30 | 45 |
| K7 | Double knit jersey, salmon | - | 11 | 85.8% | 50.2% | 15.0% | 14 | 63 | 173 |
| | | Launder | 11 | 87.9% | 71.2% | 54.6% | 12 | 31 | 55 |
| K8 | Fine-gauge, single knit jersey, 5% lycra | - | 44 | 82.0% | 42.1% | 6.4% | 5 | 20 | 62 |
| | | Launder | 21 | 89.1% | 73.1% | 44.2% | 6 | 15 | 39 |
| K10 | Spandex polyester | - | 8 | 95.4% | 83.0% | 68.8% | 6 | 23 | 47 |
| | | Launder | 7 | 89.0% | 77.3% | 55.7% | 17 | 37 | 84 |
| K11 | Nylon | - | 5 | 96.0% | 92.7% | 86.5% | 8 | 15 | 29 |
| | | Launder | 6 | 94.8% | 88.4% | 77.2% | 9 | 21 | 43 |
| CP1 | Velour | - | 8 | 88.6% | 78.0% | 58.5% | 15 | 31 | 67 |
| | | Launder | 13 | 88.0% | 75.4% | 51.1% | 10 | 22 | 52 |
| CP2 | Fleece | - | 25 | 80.3% | 48.1% | 21.6% | 9 | 29 | 61 |
| | | Launder | 16 | 87.6% | 66.5% | 35.8% | 8 | 26 | 64 |
| nW3 | H400 | - | 29 | 48.0% | 13.0% | 5.8% | 25 | 70 | 98 |
| | | Heat x 10 | 28 | 33.4% | 9.3% | 2.1% | 39 | 85 | 139 |
| | | Launder | 31 | 56.9% | 22.4% | 2.9% | 18 | 48 | 115 |
| | | Launder | 31 | 58.5% | 21.5% | 4.2% | 17 | 50 | 102 |
| | | Heat | 28 | 41.0% | 10.0% | 2.6% | 32 | 82 | 130 |
| | | IPA | 34 | 74.3% | 30.7% | 6.4% | 9 | 35 | 81 |
| | | IPA | 38 | 81.7% | 19.8% | 0.7% | 5 | 43 | 129 |
| | | SW | 41 | 68.0% | 22.0% | 4.8% | 9 | 37 | 74 |
| | | SW | 42 | 66.0% | 20.5% | 3.0% | 10 | 38 | 83 |
| | | SW | 37 | 73.1% | 28.3% | 5.3% | 8 | 34 | 79 |
| nW4 | H600 | - | 38 | 63.0% | 11.9% | 1.5% | 12 | 56 | 111 |
| | | Heat x 10 | 38 | 25.9% | 5.3% | 0.7% | 36 | 77 | 131 |
| | | Launder | 42 | 65.1% | 24.4% | 2.8% | 10 | 34 | 85 |
| | | Heat | 42 | 44.0% | 7.0% | 0.9% | 20 | 63 | 112 |
| | | IPA | - | 35.7% | 3.0% | 1.0% | - | - | - |
| | | SW | 40 | 74.0% | 27.0% | 5.6% | 8 | 33 | 72 |
| | | SW | 34 | 77.0% | 33.1% | 6.3% | 8 | 33 | 81 |
| | | SW | 29 | 75.3% | 30.0% | 6.7% | 10 | 41 | 93 |
| nW5 | Dried baby wipe | - | 5 | 89.0% | 69.0% | 53.0% | 23 | 74 | 127 |
| | | Launder | 6 | 84.8% | 67.0% | 51.3% | 27 | 67 | 111 |
| | | Heat | 5 | 90.7% | 74.2% | 55.9% | 20 | 60 | 116 |
| | | IPA | 7 | 93.9% | 75.5% | 50.1% | 9 | 40 | 99 |
| | | SW | 8 | 87.4% | 60.9% | 20.9% | 17 | 62 | 196 |
| | | SW | 8 | 91.1% | 66.1% | 24.2% | 12 | 52 | 177 |
| ASTM2 | Surgical mask | - | 21 | 0.4% | 0.0% | 0.0% | 257 | 386 | ~∞ |
| | | Old | 27 | 78.6% | 26.6% | 3.9% | 9 | 49 | 121 |
| N95 | N95, 3M 1860 | - | 41 | 0.0% | 0.0% | 0.0% | ~∞ | ~∞ | ~∞ |
| | | IPA | 45 | 12.0% | 1.6% | 0.0% | 47 | 92 | ~∞ |
| | | IPA | 46 | 13.0% | 1.8% | 0.0% | 44 | 87 | ~∞ |
| | | IPA | 47 | 13.0% | 1.8% | 0.6% | 43 | 85 | 109 |
| | | IPA | 46 | 13.7% | 1.7% | 0.2% | 43 | 89 | 136 |
| | | Heat | 53 | 0.0% | 0.0% | 0.0% | ~∞ | ~∞ | ~∞ |

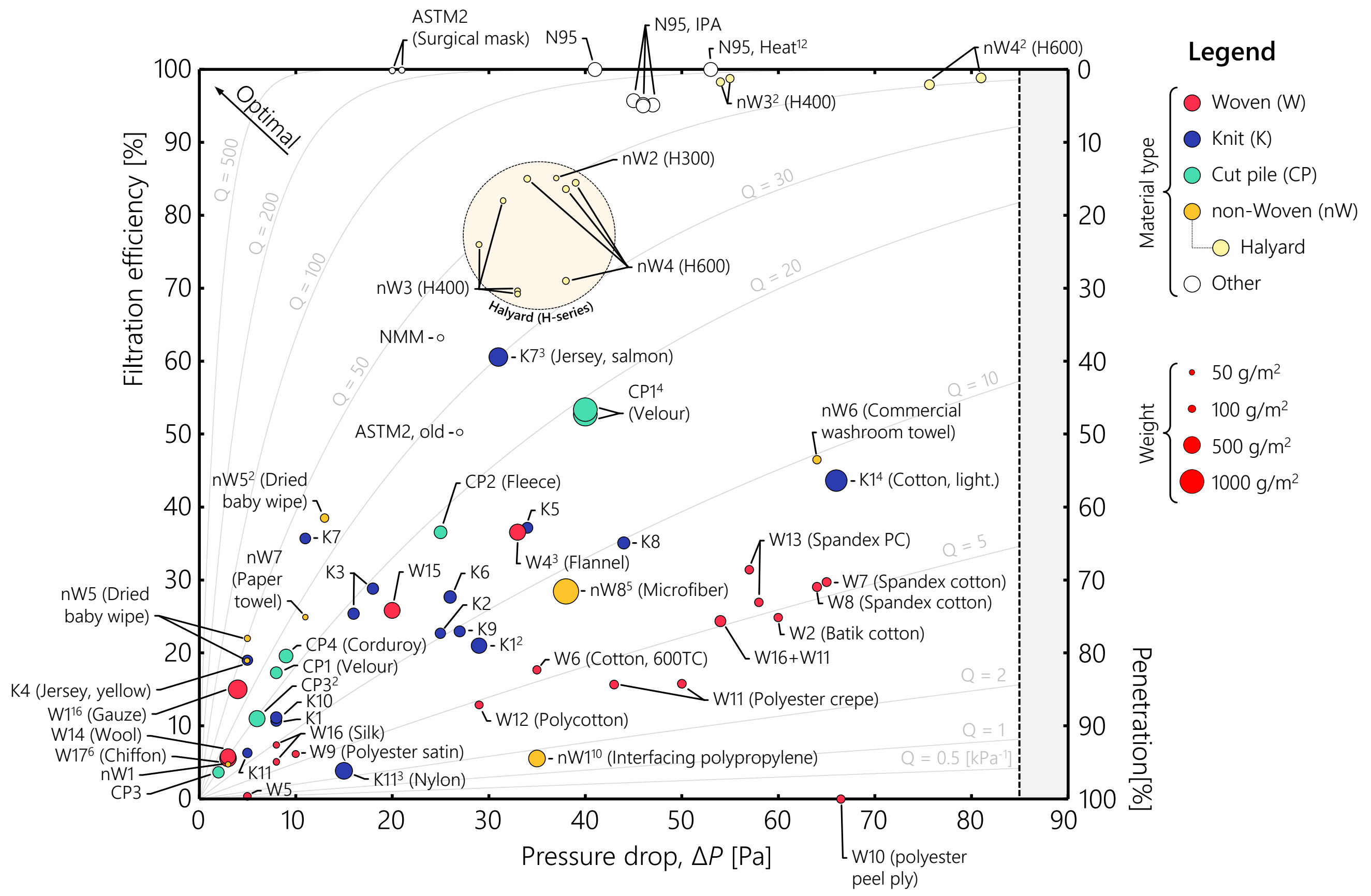

**Figure S1.** An analogous plot Error! Reference source not found. for 1.9 micron aerodynamic diameter particles, showing pressure drop and material penetration for a face velocity of 4.9 cm/s and a range of common materials. Quality factor isolines are labelled in kPa$^{-1}$. In general, penetrations are higher (i.e. filtration efficiencies are lower) for the smaller particle size considered here.